# Scalable, Non-Invasive Glucose Sensor Based on Boronic Acid Functionalized Carbon Nanotube Transistors

Mitchell B. Lerner[1], Nicholas Kybert[1], Ryan Mendoza[2,*], Romain Villechenon[1,†], Manuel A. Bonilla Lopez[1,‡], and A.T. Charlie Johnson[1,2§]

1 Department of Physics and Astronomy, University of Pennsylvania, Philadelphia, PA 19104, USA
2 Department of Electrical and Systems Engineering, University of Pennsylvania, Philadelphia, PA 19104, USA

We developed a scalable, label-free all-electronic sensor for D-glucose based on a carbon nanotube transistor functionalized with pyrene-1-boronic acid. This sensor responds to glucose in the range 1μM – 100 mM, which includes typical glucose concentrations in human blood and saliva. Control experiments establish that functionalization with the boronic acid provides high sensitivity and selectivity for glucose. The devices show better sensitivity than commercial blood glucose meters and could represent a general strategy to bloodless glucose monitoring by detecting low concentrations of glucose in saliva.

---

[*] Present address: Graphene Frontiers, Philadelphia, PA 19104, USA
[†] Present address: Phelma, Grenoble Institute of Technology, Grenoble 38016, France
[‡] Present address: Department of Physics and Electronics, University of Puerto Rico at Humacao, Humacao, Puerto Rico 00792, USA
[§] Correspondence to: A.T. Charlie Johnson. Email: cjohnson@physics.upenn.edu

Carbon nanotube field effect transistors (NT FETs) provide a unique platform for biosensing applications.[1] Since every atom is on the surface, carbon nanotubes are highly sensitive to small changes in their immediate surroundings, making them ideal readout elements for chemical sensors. Ease of fabrication, well-understood carbon chemistry, and fast electronic readout ($\leq 1$ ms) make functionalized NT FETs desirable as chemical sensors for biomolecular detection.[2]

Diabetes mellitus affects nearly 300 million people worldwide and its incidence is expected to increase rapidly in the coming decades.[3] Continuous and accurate monitoring of patient blood glucose levels is critical for diagnosis and management of the disease. A powerful approach for detecting glucose in fluid is complexation by boronic acid moieties.[4] This method is superior to the more common enzymatic detection strategies because it is not affected by factors that affect mass transport of the analyte and enzyme activity.[5] Commercially available glucose sensors consume the analyte, must prevent protein denaturation over a long time period, and require mediators to transport electrons into the conduction channel.[6,7] In contrast, boronic acid-mediated detection is based on equilibrium thermodynamics and does not require special treatment of the sensor to maintain structural integrity.[7] Complexation of boronic acid with a monosaccharide results in a boronate anion, which can affect the local electrostatic environment surrounding the nanotube.[8] Nanotube-based glucose sensors have been demonstrated using a variety of techniques;[9] however, the sensitivity of nanotube FETs to the local charge environment suggests that this modality could potentially attain the lowest detection limits.[10,11] The prevalence of boronate anions created from bound glucose molecules is expected to modulate the electronic transport properties of the FET in a concentration-dependent manner. Carbon nanotube-based sensors with sensitivity equal to or greater than commercially available enzyme-based immunoassays would then potentially be useful for glucose monitoring in bodily fluids other than blood, such as saliva, eliminating the need for daily, uncomfortable finger pricking.

Biosensors that combine chemical elements for analyte recognition with an all-electronic, nano-enabled readout element would be ideal for medical diagnosis if they could be made cost effective. The use of solutions enriched in semiconducting nanotubes[12] for wafer-scale fabrication of thin-film transistors has been reported,[13] and this approach was recently extended in a demonstration of the production of large arrays of reproducible vapor sensors based on DNA-functionalized NT transistors.[14] A similar process was used to create the devices used in these experiments, offering the prospect that they could be produced with high yield.

Experiments were performed using solutions of carbon nanotubes comprised of 98% semiconducting NTs (NanoIntegris, Inc.), with a concentration of 10 mg/L. Transistor arrays were fabricated on oxidized silicon wafers (oxide thickness 500 nm). Atomic layer deposition (ALD) was used to create a monolayer of aminopropyltriethoxysilane (APTES) to promote adhesion of the carbon nanotubes,[15] which can also improve the performance of nanotube-based sensors[16]. Drops of NT solution were pipetted onto the APTES-coated wafer and incubated in a humid environment for 20 minutes. The wafer was then washed in two successive isopropanol baths for 5 minutes each, followed by a deionized water bath for 10 minutes, and the wafer was then dried using compressed nitrogen gas. Source and drain electrodes were composed of 50 nm Ti defined via a photolithographic process optimized for NT devices[17] and metallized in an electron beam evaporator. Titanium was chosen purposefully as the electrode material because it forms a self-passivating oxide layer under ambient, thus ensuring that any sensing response is due to the interactions between the analyte and the NTs rather than the metal contacts.[18] This completed the three-terminal field effect transistor (FET) geometry with the silicon backplane acting as a global backgate. Typical device dimensions were 10 μm channel length and 15 μm channel width. The NT density was approximately 1-2/μm$^2$, yielding 150-300 NTs in the channel region that formed several conducting paths connecting source and drain.[19]

The efficacy of functionalization of NT transistors with pyrene compounds, including pyrene-1-boronic acid, has been documented previously by us via atomic force microscopy, X-ray photoelectron spectroscopy, and electronic measurements.[10] FET devices were submerged in 5 μM pyrene-1-boronic acid in acetonitrile for 2 hours, then transferred to baths of acetonitrile, isopropanol, and DI water for 5 minutes each and dried with compressed nitrogen. The sample was then baked on a hot plate at 120°C for 2 hours in order to drive off any remaining solvents.

Saccharide solutions (glucose, lactose) were prepared by dissolving a known mass of sugar in DI water to achieve the desired concentrations. To start a sensing experiment, two 500 nL droplets of pure DI water were pipetted onto the channel region to allow the sensor to equilibrate to liquid temperature, pH, ionic strength, etc. Glucose/lactose solutions of increasing concentrations were then added to the existing water droplets in 500 nL increments with the resulting equilibrium concentrations used in measuring sensor responses. Each glucose/lactose concentration was measured on 6-8 different devices to ensure reproducibility. Measurements were conducted in a humid environment with low evaporation rate (< 10% of the total drop volume over the course of the entire experiment). The principle of operation is illustrated in Figure 1.

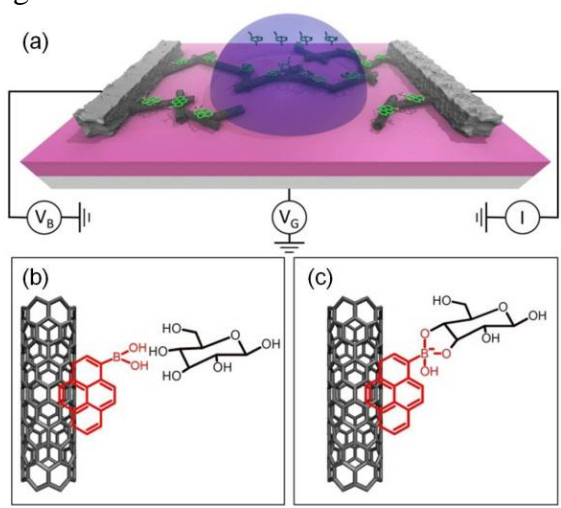

**Figure 1. a) Schematic of the experimental geometry. b) Illustration of glucose binding to a nanotube functionalized with pyrene-1-boronic acid. c) Bound glucose forms a boronate anion complex that has electrostatic effects on the nanotube FET.**

Figure 2 shows current-gate voltage (I-$V_g$) characteristics from the as-fabricated NT FET, after functionalization with pyrene-1-boronic acid, and after exposure to glucose solution. Following incubation in pyrene-1-boronic acid, devices showed an increase of ~ 2-3V in threshold voltage $V_{TH}$, defined as the voltage required to deplete carriers in the channel. The threshold voltage was calculated by fitting the I-$V_g$ curve with a tangent line at the point of maximum transconductance and determining the intercept of this line with the gate voltage axis[20]. This increase in $V_{TH}$ can be ascribed to electrostatic "chemical gating" of the device by negatively charged boronic acid moieties in the presence of a thin water layer that accumulates on the hydrophilic substrate.[10] Functional groups on the substrate surface can change the local pH in this thin water layer by as much as two pH units.[21] The proton-accepting, amine-terminated functional groups in the APTES monolayer would be expected to raise the pH in the interfacial water layer, making it slightly basic. Under these conditions, the pyrene-1-boronic acid with a pK$_a$ of 8.8[22] would be significantly deprotonated (13.6% deprotonated for pH 8 and 31.9% for pH 8.5), so each molecule acts as a negatively charged local electrostatic gate. In order to compensate for the effect of this local gate, the field supplied by the global backgate would have to be more positive, hence the shift to the right in the threshold voltage on the I-V$_g$ characteristic. Upon introduction of glucose, the threshold voltage increased slightly (1-2 V depending on glucose concentration) and the ON state current (current at V$_g$ = -10V) decreased by a factor that varied systematically with glucose concentration as detailed in the Supplemental Material.

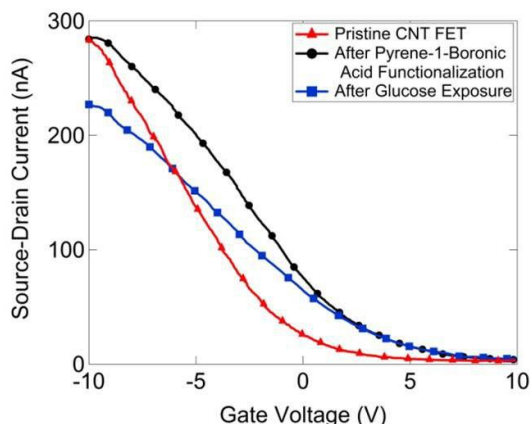

**Figure 2. Current-gate voltage characteristic from a nanotube FET after each step in the experiment. Data from the as-fabricated nanotube FET is shown in red with triangular markers. The measured curve after functionalization with pyrene-1-boronic acid is shown in black with circle markers. The measured curve after exposure to a 1mM glucose solution is shown in blue with square markers. Bias voltage is 100 mV.**

Measurements of sensor response as a function of glucose concentration were taken with the back gate voltage set to -5V, which was a region of maximum transconductance for all samples used. To account for differences in device resistance, the sensor response was recorded as the fractional change in DC current at constant bias voltage ($\Delta I/I$). Device resistance was monitored as successive drops of glucose solution were added to the channel region in order to progressively increase the overall glucose concentration. It was observed that the fractional decrease in device current grew monotonically as the concentration of glucose was increased. Moreover, this fractional decrease in current was found to depend only on the glucose concentration and was independent of the sequence of glucose concentrations delivered to the sensor and the time between droplet deliveries (Fig. 3). Expected response times were informed by a calculation of the diffusion limits for glucose molecules in water. The diffusion length is given by $L = \sqrt{Dt}$,[23] where $D$ is the diffusion constant (6.7 x 10$^{-6}$ cm$^2$/s for glucose in water[24]) and $t$ is the diffusion time. Analyte droplets were observed to spread over the hydrophilic SiO$_2$/APTES substrate such that droplet height did not exceed ca. 60 µm. From the diffusion equation, this corresponds to a diffusion time of 1.3 sec, which is the estimated upper bound for the equilibration time for sensor within the drop. Sensors were found to respond rapidly, in rough agreement with this estimate. Sensor recovery was significantly slower, consistent with the fact that a covalent bond was formed between and the boronic acid receptor. It was found that the sensor returned to baseline after ca. 1 hour (2 minutes) in static (flowing) DI water at room temperature or about 10 min in static DI water at 50 °C.

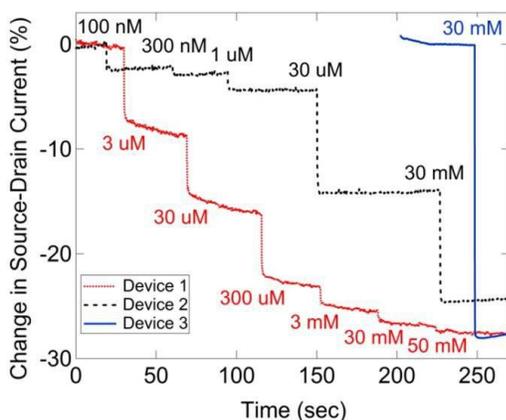

**Figure 3.** Normalized reduction in source-drain current for three devices. Regardless of the sequence of droplets added to the sensor, a glucose concentration of 30 mM produces a response of approximately 26% ± 2%.

From a collection of time traces similar to those in Fig. 3, the sensor response as a function of glucose concentration was calculated. Each concentration was tested on 5-8 devices to ensure reproducibility. The sensor responses are shown in Fig. 4; the error bars are the standard error of the mean. The data agree with a model based on the Hill-Langmuir equation for equilibrium ligand-receptor binding (Fig. 4):[25]

$$\frac{\Delta I}{I} = A \frac{\left(c/K_d\right)^n}{1+\left(c/K_d\right)^n} + Z$$

Here $c$ is the glucose concentration, $A$ is the sensor response at saturation when all glucose binding sites are occupied, $Z$ is an offset that accounts for the response to pure water, $K_d$ is the dissociation constant describing the concentration at which half of available binding sites are occupied, and $n$ is the Hill coefficient describing cooperativity of binding.

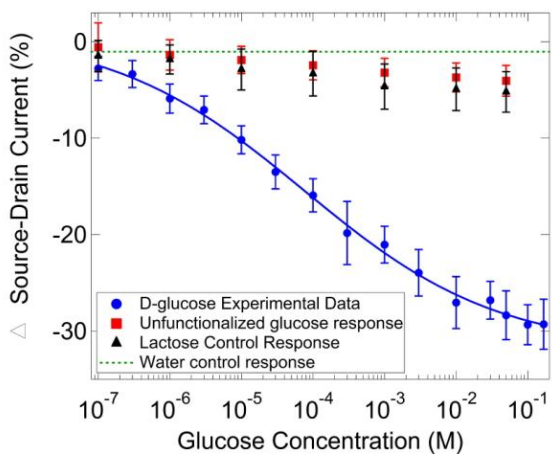

**Figure 4.** Sensor response as a function of glucose concentration (blue circles) with the corresponding Hill-Langmuir fit (blue line). Also included are data from three control measurements. In the first, un-functionalized devices show minimal response to glucose exposure (red squares). In the second, devices functionalized with pyrene-1-boronic acid show minimal response upon exposure to lactose solution, used as a negative control (black triangles). Functionalized devices show a null response upon exposure to DI water (green dashed line).

The best fit to the glucose data yielded a saturated response $A = -31.0 \pm 2$ %, offset parameter $Z = -1.0 \pm 1.5$%, dissociation constant $K_d = 76$ µM $\pm 8.2$ µM, and $n = 0.32 \pm 0.04$. The best fit value of the offset parameter $Z = -1.0 \pm 1.5$% was statistically indistinguishable from the experimentally measured responses of seven devices to DI water as a negative control ($\Delta I/I = -0.50\% \pm 0.54\%$). The value of the dissociation constant determined from the fit (76 µM $\pm 8.2$ µM) describes the concentration of glucose at which half the receptors are occupied and the sensor has its maximum differential response. This regime coincides with glucose levels of diagnostic significance. For example, Jurysta et al. used a complex, enzymatic assay and found a statistically significant difference between salivary glucose concentrations for diabetic and non-diabetic patients: $78.7 \pm 9.2$ µM and $201.9 \pm 34.9$ µM, respectively.[26] This variation could potentially be detected in real time with a refined sensor system based on the methodology presented here.

The best fit value of the cooperativity parameter, $n = 0.32 \pm 0.04$, indicates negatively cooperative binding of glucose to the boronic acid receptor in the context of the NT FET biosensor. The interpretation is that binding of two glucose molecules by adjacent boronic acids is suppressed, possibly because the boronate anion formed when the first glucose molecule binds to a boronic acid creates an electrostatic radius within which additional bound anions are unstable. The data presented in Fig. 4 show that the measured responses from a collection of 6-8 devices could be used to discriminate between pure water ($\Delta I/I = -0.50\% \pm 0.54\%$) and water containing glucose at a concentration of 300 nM ($\Delta I/I = -3.31\% \pm 1.30\%$).

The observed reduction in source-drain current is consistent with increased carrier scattering due to the presence of boronate anions formed by bound glucose in the local electrostatic environment of the carbon nanotube. As described in the Supplemental Material, a single device exposed to increasing concentrations of glucose solution showed a monotonic decrease in transistor mobility, consistent with carrier scattering being responsible for the decrease in device conduction. As a control experiment, boronic acid-functionalized sensors were exposed to a solution of lactose in the same manner as glucose exposure. Lactose, a 1→4 oligosaccharide, is known to have a low binding affinity for boronic acid moieties.[27] Figure 4 shows the sensor response to several concentrations of lactose. Exposure to lactose at concentrations below 1 mM was not distinguishable from the response to pure water, and the signal saturates at a low level of approximately -6%. This small decrease in current is likely due to non-specifically bound molecules increasing carrier scattering at these high sugar concentrations. In a second control experiment, unfunctionalized devices were exposed to glucose solutions. The average responses, calculated using 8 devices, were below 5% even for glucose concentrations as high as 10 mM. We thus concluded that the sensing response from boronic acid-functionalized carbon nanotube transistors is specific for glucose, and that the pyrene chemistry is required to effectively bind glucose with high density along the nanotube sidewall.

To summarize, we developed a robust and reproducible fabrication method for carbon nanotube-based glucose sensors that exhibited excellent sensitivity and selectivity. Device responses were shown to vary systematically with glucose concentration in a clinically relevant range, with a minimum detection limit of 300 nM. Control experiments were used to confirm the effectiveness of the functionalization chemistry and the selectivity of the sensor for glucose over another sugar. Potential applications include glucose sensors for diabetics based on either blood or salivary glucose levels where the pyrene boronic acid functionalized nanotube FET provides chemical specificity and all-electronic readout. The incorporation of solution phase nanotubes makes such devices promising candidates for scalable, point-of-care diagnostic tools. Such devices would potentially eliminate the need for frequent, uncomfortable finger pricking for blood glucose measurements and would greatly improve diabetic peoples' quality of life while still maintaining a high level of diagnostic accuracy.

**Acknowledgements**: This work was supported by the Nano/Bio Interface Center through the National Science Foundation NSEC DMR08-32802. Use of the facilities of the Nano/Bio Interface Center is also acknowledged. Mitchell Lerner acknowledges the support of a SMART Fellowship, and Manuel Lopez was supported by NSF REU site grant DMR-1062638.